\documentclass[a4paper,11pt]{article}
\usepackage{pos}

\usepackage{xcolor}

\usepackage{braket}
\usepackage{graphicx}
\usepackage{amsmath}
\usepackage{extarrows}

\newcommand{\avgxxx}{\langle x^3 \rangle}

\newcommand{\rr}{\langle r^2 \rangle}

\newcommand{\be}{\begin{equation}}
\newcommand{\ee}{\end{equation}}
\newcommand{\bea}{\begin{eqnarray}}
\newcommand{\eea}{\end{eqnarray}}

\newcommand{\Op}{{\cal O}}
\newcommand{\avgx}{\langle x \rangle}

\title{Pion and kaon form factors using twisted-mass fermions}

\author[a,b]{Constantia Alexandrou}
\author[b]{Simone Bacchio}
\author[c]{Ian Clo\"et}
\author[d]{Martha Constantinou}
\author*[d]{Joseph Delmar}
\author[a,b]{Kyriakos Hadjiyiannakou}
\author[b]{Giannis Koutsou}
\author[c,d]{Colin Lauer}
\author[e]{Alejandro Vaquero Avil\'{e}s-Casco}

\affiliation[a]{Department of Physics, University of Cyprus,\\  P.O. Box 20537,  1678 Nicosia, Cyprus}

\affiliation[b]{Computation-based Science and Technology Research Center,
  The Cyprus Institute,\\ 20 Kavafi Str., Nicosia 2121, Cyprus}

\affiliation[c]{Physics Division, Argonne National Laboratory,\\ Lemont, Illinois 60439, USA}

\affiliation[d]{Department of Physics, Temple University,\\ 1925 N. 12th Street, Philadelphia, PA 19122-1801, USA}

\affiliation[e]{Department of Physics and Astronomy, University of Utah,\\ Salt Lake City, Utah 84112, USA}

\emailAdd{jdelmar@temple.edu}

\abstract{We present a calculation of the scalar, vector and tensor pion and kaon form factors using one ensemble of two degenerate light, a strange and a charm quark ($N_f=2+1+1$) of maximally twisted mass fermions with clover improvement. The quark masses are chosen so that they produce a pion mass of about 265 MeV, and a kaon mass of 530 MeV. The lattice spacing of the ensemble is 0.093 fm and the lattice has a spatial extent of 3 fm. We use a rest frame, as well as a boosted frame to obtain the form factors for a wider and denser set of four-vector momentum transfer squared, $Q^2$. To assess and eliminate excited-states contamination, we analyze several values of the source-sink time separation within the range of 1.12 -- 2.23 fm (1.12 -- 1.67 fm) for the rest (boosted) frame. The $Q^2$ dependence of the form factors is parametrized using a monopole fit, which leads to the extraction of the corresponding radius, and the tensor anomalous magnetic moment for the tensor form factor. The results for these parametrizations are compared for the pion and kaon to assess the level of the SU(3) flavor symmetry breaking.}

\FullConference{%
 The 38th International Symposium on Lattice Field Theory, LATTICE2021
  26th-30th July, 2021
  Zoom/Gather@Massachusetts Institute of Technology
}

\begin{document}

\maketitle

\section{Introduction}
Quantum chromodynamics (QCD), the theory of the strong interaction, requires a non-perturbative approach at the hadronic energy. The various form factors that can be extracted through the QCD factorization of the cross-section of physical processes give access to properties describing the structure of hadrons. The scalar form factors, for instance, can be used to explore the interplay of the emergent hadronic mass (EHM), a mechanism used to describing the large mass of hadrons~\cite{Cui:2020dlm,Roberts:2021nhw}, and the Higgs boson interaction, which increases the masses of the Goldstone bosons. The vector form factors provide insight to electromagnetic properties, and the tensor form factors are useful for beyond the standard model studies. The importance of the pion and kaon to understanding the long-range dynamics of QCD is well-established~\cite{Hagler:2009ni} through an extensive experimental investigations of the pion since the 1970s.

In this work we calculate of the scalar, vector, and tensor form factors of the pion and kaon. We consider only the connected contributions, as we expect that disconnected contributions are small, at least for the vector and tensor form factors, for larger than physical pion mass. We present the $Q^2$ dependence of the form factors and derived quantities, such as the radii and tensor anomalous magnetic moment.

\section{Theory and Lattice Setup}

We obtain the form factors for a particular flavor $f$ from the matrix elements of ultra-local operators
\begin{equation}
    \langle M({p}') | \Op^f_\Gamma | M({p}) \rangle \,,
\end{equation}
where the operator structure $\Op^f_\Gamma $ for 0-spin mesons is the scalar, $\Op^f_S=\bar{\psi}\hat{1}\psi$, vector, $\Op^f_V=\bar{\psi}\gamma^\mu\psi$, and tensor, $\Op^f_T=\bar{\psi}\sigma^{\mu\nu}\psi$ with  $\sigma^{\mu\nu}=\frac{1}{2}[\gamma^\mu,\gamma^\nu]$. We extract the 4-vector momentum transfer, $t\equiv-Q^2$, dependence of the form factor from the off-forward matrix element, where the momentum transfer between the initial (${p}$) and final (${p'}$) state is ${Q}={p'} - {p}$. The decomposition of each matrix element for the general frame in Euclidean space is~\cite{Hagler:2009ni}
\begin{align}
    \langle M({p}') | \Op^f_S | M({p}) \rangle &= \frac{1}{\sqrt{4 E(p) E(p')}} A^{M^f}_{S10}\,, \\[1ex]
    \langle M({p}') | \Op^f_{V^\mu} | M({p}) \rangle &= -i\, \frac{2\, P^\mu}{\sqrt{4 E(p) E(p')}} \, A^{M^f}_{10}\,, \\[1ex]
    \langle M({p}') | \Op^f_{T^{\mu\nu}} | M({p}) \rangle &= i\, \frac{(P^\mu \Delta^\nu - P^\nu \Delta^\mu)}{m_M \sqrt{4 E(p) E(p')}}  \,B^{M^f}_{T10}\,.
        \label{eq:tensor_decomp2}
\end{align}
$P^\mu$ is the average momentum, $P \equiv (p'+p)/2$, and $\Delta$ is the momentum difference, $\Delta \equiv p'-p$. The mass of meson $M$ is indicated by $m_M$, and its energy at momentum $\vec{p}$ is $E(p){=}\sqrt{m_M^2 + \vec{p}\,^2}$. We omit the index $M$ from the energy to simplify the notation. Here we use the notation $F^{M,f}_S \equiv A^{M^f}_{S10}$, $F^{M,f}_V \equiv A^{M^f}_{10}$,  $F^{M,f}_T \equiv B^{M^f}_{T10}$. 

We use an ensemble of twisted-mass clover fermions and Iwasaki improved gluons with pion mass 265 MeV and $a=0.09471(39)$ fm.  The ensemble contains the two light mass-degenerate quarks as well as the strange and charm quarks in the sea ($N_f=2+1+1$). Additionally, the volume ($L^3\times T$) is $32^3\times64$, $L m_\pi=4$, and $L=3.0$ fm. These gauge configurations have been produced by the Extended Twisted Mass Collaboration (ETMC)~\cite{Alexandrou:2018egz}. 

We extract matrix elements in both the rest and boosted frames. For the latter, we choose a momentum transfer of the form $\mathbf{p'}=2\pi \mathbf{n'}/L$ with $\mathbf{n'}=(\pm1,\pm1,\pm1)$. This choice is such that one can extract matrix elements with up to three covariant operators~\cite{Alexandrou:2021mmi}, avoiding any mixing under renormalization. The fact that we have eight combinations of the momentum boost increases the computational cost by a factor of eight. We note that, the lattice data these combinations can be averaged in the forward limit, as we have done for $\avgx$ - $\avgxxx$~\cite{Alexandrou:2020gxs,Alexandrou:2021mmi}. However, this does not apply for the form factors because the various $\mathbf{p'}$ do not correspond to the same value of $Q^2$ in the boosted frame. 

The statistics for the rest frame is the same for all values of the source-sink time separations ($t_s/a=12,14,16,18,20,24$) and equal to 1,952. For the boosted frame we have a statistics of 46,848 for $t_s/a=12$, and 101,504 for $t_s/a=14,16,18$. More details can be found in Ref.~\cite{Alexandrou:2021ztx}.

We extract the meson matrix moments using the optimized ratio
\begin{equation}
    \label{eq:ratio}
    R^M_\Gamma(t_s,t;\mathbf{p}', \mathbf{q} = \mathbf{p}' - \mathbf{p}) = \frac{ C^M_\Gamma(t_s,t;\mathbf{p}',\mathbf{q}) }{ C_M(t;\mathbf{p}'^2) }
    \sqrt{ \frac{ C_M(t_s-t;\mathbf{p}^2) C_M(t;\mathbf{p}'^2) C_M(;t_s\mathbf{p}'^2) }{ C_M(t_s-t;\mathbf{p}'^2) C_M(t;\mathbf{p}^2) C_M(t_s;\mathbf{p}^2) } } \,,
\end{equation}
which cancels the time dependence in the exponentials, as well as the overlaps between the interpolating field and the meson state. As the ratio in Eq.~\eqref{eq:ratio} is written for a general frame, we use $C_M(t;\mathbf{p}^2)=c_0 e^{-E_0(\mathbf{p}^2) t}$ for the two-point functions, where $c_0$ is calculated from the two-state fit on the two-point functions, and $E_0=\sqrt{m^2 + \mathbf{p}^2}$ is calculated from the plateau fit on the effective mass. At insertion times far enough from the source and sink positions, the ratio becomes independent of insertion time,
\begin{equation}
    R^M_\Gamma(t_s,t;\mathbf{p}', \mathbf{q}) \xlongrightarrow[\text{$Et\gg1$}]{\text{$\Delta E(t_s-t)\gg1$}}  \Pi^M_\Gamma(t_s; \mathbf{p}', \mathbf{q}) \,.
    \label{eq:ratio2}
\end{equation}
We use two methods to calculate $\Pi^M_\Gamma$: \textbf{(a)} by fitting the plateau region of the data to a constant value; \textbf{(b)} by performing a two-state fit on the three-point functions. Combining these methods, we can study and eliminate excited-states contamination. Representative results are shown in the next section.

\section{Results on Form Factor}

\begin{figure}[h!]
\begin{minipage}{6cm}
    \includegraphics[scale=0.20]{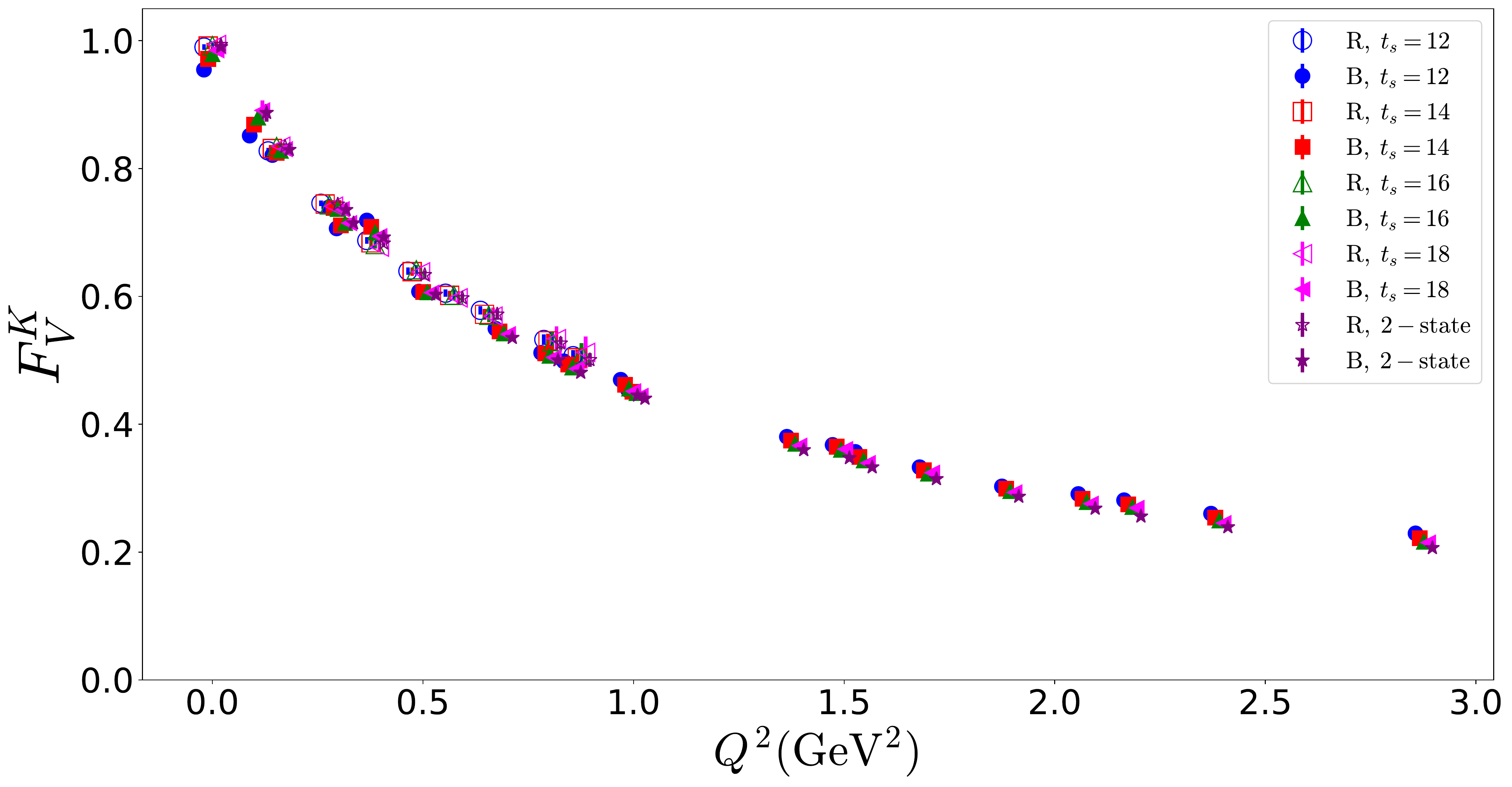}
\end{minipage}
\hfill\begin{minipage}{6cm}
    \caption{Comparison of the vector form factor of the kaon between the rest (open symbols) and boosted frame (filled symbols). The two-state fit as applied on these data is shown with purple stars. Blue, red, green, and magenta points correspond to $t_s/a=12,\,14,\,16,\,18$. Statistical errors are included, but are too small to be visible.}
    \label{fig:FK_v_all}
    \end{minipage}
\end{figure}

\begin{figure}[h!]
    \centering
     \includegraphics[scale=0.21]{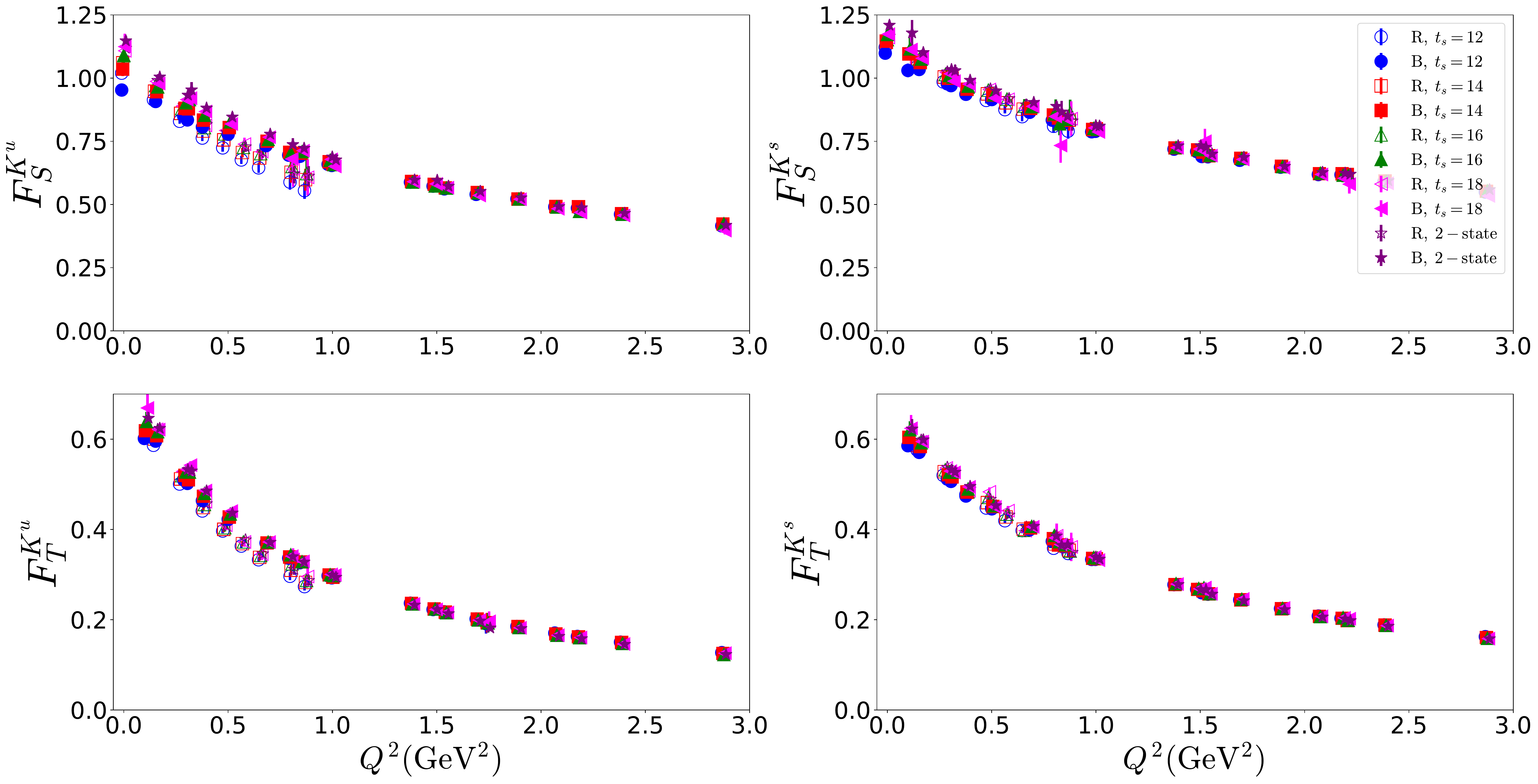}
    \caption{Comparison of the scalar and tensor form factor of the kaon between the rest and boosted frame. The notation is the same as Fig.~\ref{fig:FK_v_all}. Statistical errors are included, but are too small to be visible. }
    \label{fig:FK_s_t_all}
\end{figure}

Due to space limitations, we only show selected results on the kaon form factors. The complete set of results can be found in Ref.~\cite{Alexandrou:2021ztx}. In Fig.~\ref{fig:FK_v_all} we show a comparison between the rest and boosted frame for the vector form factor, and in Fig.~\ref{fig:FK_s_t_all} for the up and strange contributions to the scalar and tensor ones. We only include $t_s/a=12,14,16,18$ for better visibility, as well, the two-state fits. We find that the results for the vector become fully compatible between the two frames at $t_s$ values where excited-state are eliminated. There is some tension in the slope between the two frames for the up-quark scalar and tensor form factors, with the rest-frame results being a bit lower. The strange-quark scalar and tensor form factors are compatible for the two frames.

\vspace*{-0.40cm}

\section{Parametrization of Form Factors}

We parameterize the $Q^2$ dependence of the form factors using the monopole ansatz depicted by the Vector Meson Dominance (VMD) model~\cite{OConnell:1995fwv},
\begin{equation}
\label{eq:fit}
F_\Gamma(Q^2) = \frac{F_\Gamma(0)}{1 + \frac{Q^2}{{\cal M}_\Gamma^2}} \,,
\qquad\qquad
\langle r^2 \rangle_\Gamma = -\frac{6}{F_\Gamma(0)} \frac{\partial F_\Gamma (Q^2)}{\partial Q^2}\Bigg{|}_{Q^2=0} =  \frac{6}{M^2_\Gamma}\,.
\end{equation}
$F_\Gamma(0)$ is the forward limit of the form factor, and ${\cal M}_\Gamma$ is the monopole mass. For the scalar and vector form factors, we also employ a one-parameter fit by fixing $F_\Gamma(0)$ to the value obtained from our lattice data. The radius, $\langle r^2 \rangle_\Gamma$, is an interesting quantity that is defined as the slope of the form factor at $Q^2=0$ and can be obtained from ${\cal M}_\Gamma$ as shown above.

For the parameterization, we utilize the results from the two-state fits to ensure that excited-states are eliminated. We apply the fit of Eq.~(\ref{eq:fit}) to the results of the rest frame, the boosted frame, and a combination of both frames. For the pion, we test the $Q^2$ values up to 0.55, 1, and 2.5 GeV$^2$. For the kaon, we test $Q^2$ up to 1 and 3 GeV$^2$. In the case of both mesons, we choose the values of $F_\Gamma$ and $M_\Gamma$ from the combined fit and the entire $Q^2$ range. 

\begin{figure}[h!]
\begin{minipage}{7cm}
    \includegraphics[scale=0.25]{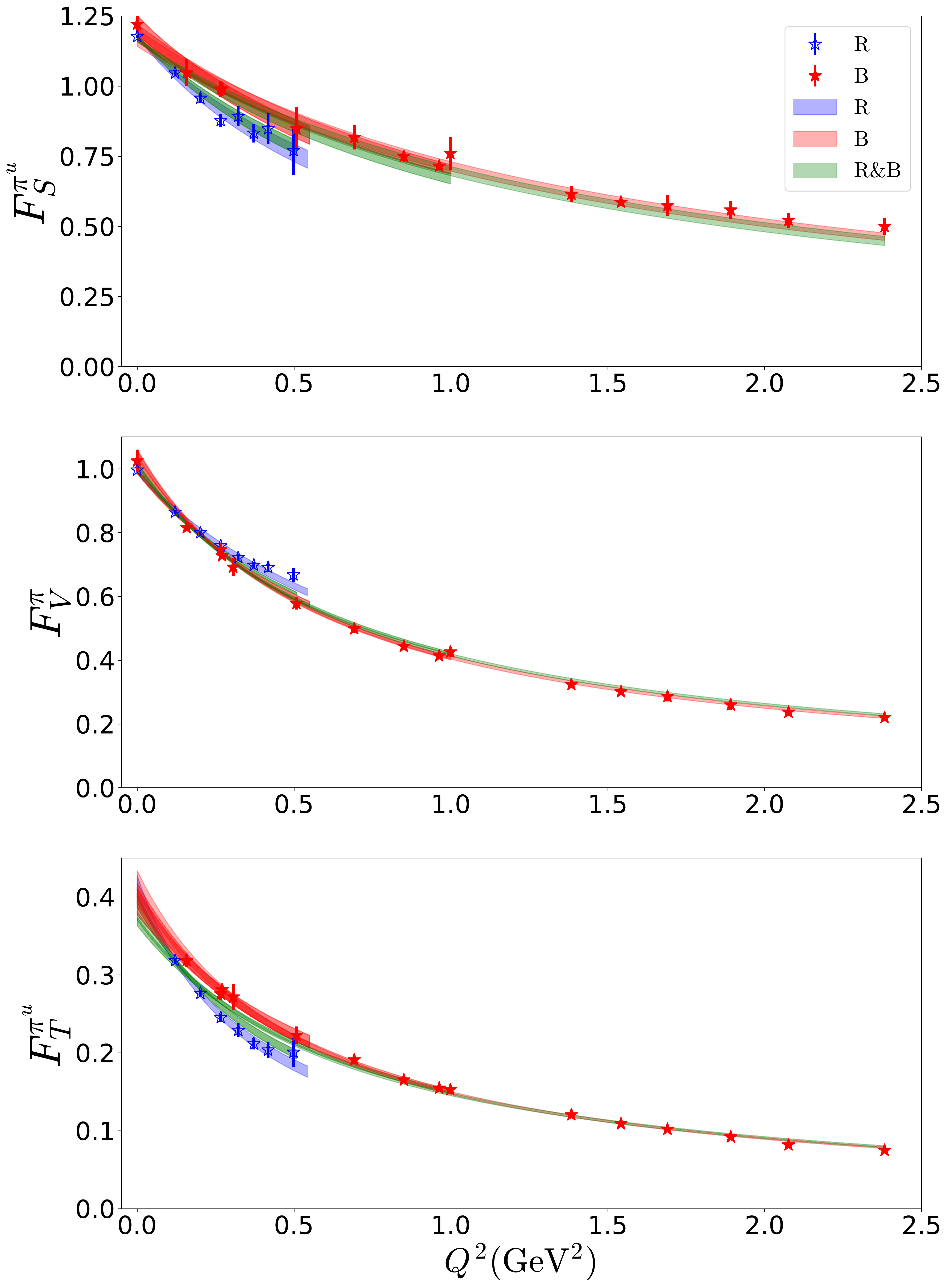}
    \end{minipage}
\hfill\begin{minipage}{5cm}
    \caption{From top to bottom: The scalar, vector and tensor form factors of the pion using two-state fit in the rest (blue points) and boosted (red points) frame. The two-parameter fitted form factors are shown with bands for the cases of the rest frame R  (blue),  boosted frame B (red), and combined data R$\&$B (green). The length of the band indicates the at $Q_{\rm max}^2$ interval used for the fit. Statistical errors are included, but are too small to be visible. }
    \label{fig:Pion_fit}
\end{minipage}
\end{figure}

In Fig.~\ref{fig:Pion_fit} we plot $F_\Gamma(Q^2)$ using the two-state fit data in the rest and boosted frames for the pion. We compare these against the fitted form factors for the cases described above. There is a small difference between the fits of the rest and boosted data sets and the two-state fit data for the case of $Q_{\rm max}^s=0.55$ GeV$^2$. The fits of the combined data sets fall between those of the individual rest and boosted fits, as is expected. We also find agreement between the bands of  $F_S$ and $F_V$ and the corresponding value at $Q^2=0$. The discrepancy for the case of $\kappa_T$ discussed is due to the change in the slope for different data sets.
    
\begin{figure}[h!]
    \centering
    \includegraphics[scale=0.25]{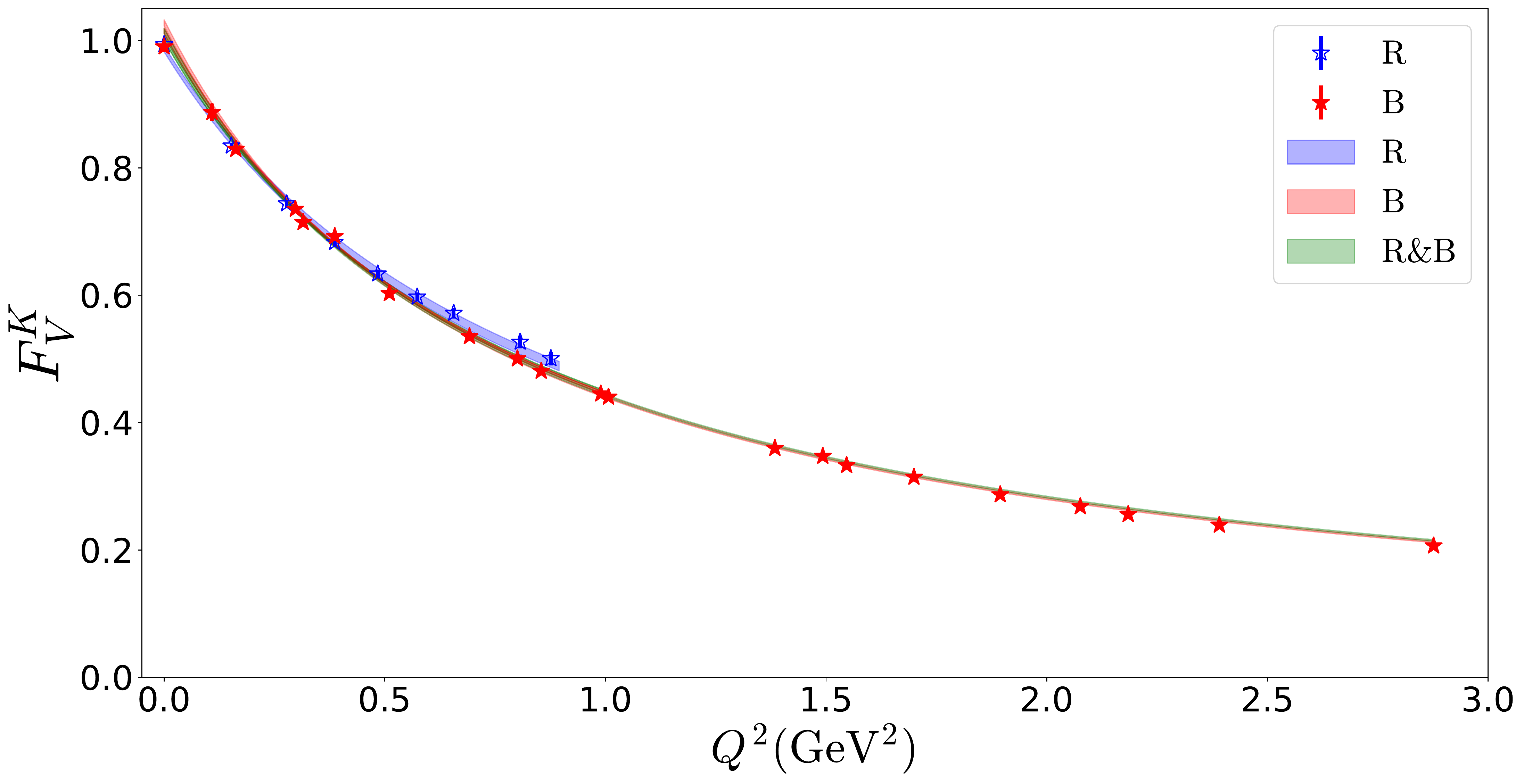}
    \caption{Parametrization of the vector form factor for the kaon. The notation is the same as Fig.~\ref{fig:Pion_fit}.}
    \label{fig:kaon_fit_v}
\end{figure}

\begin{figure}[h!]
    \centering
    \includegraphics[scale=0.23]{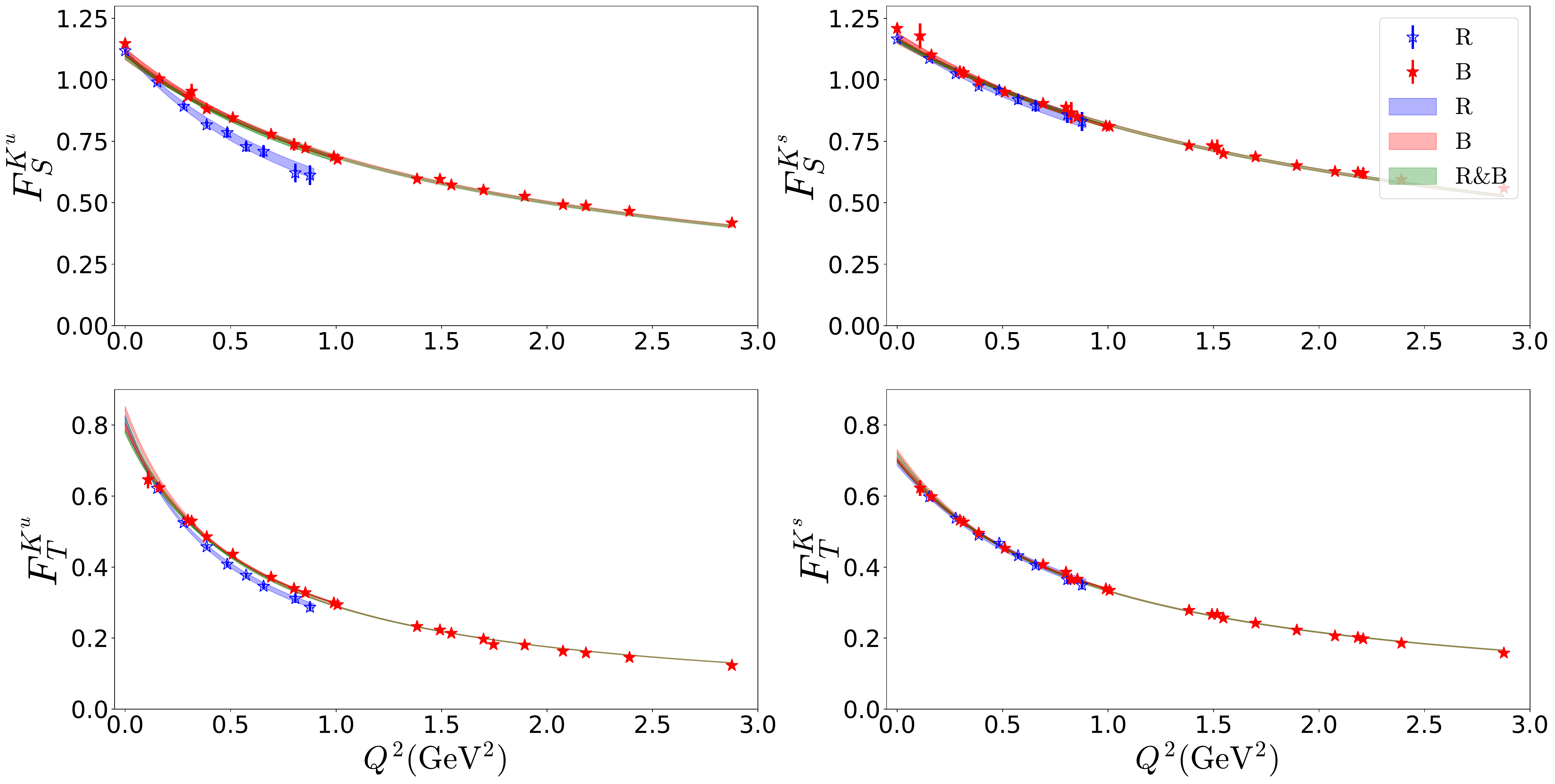}
    \caption{Parametrization of the scalar and tensor form factor for the up (left) and strange (right) quark components of the kaon. The notation is the same as Fig.~\ref{fig:Pion_fit}. }
    \label{fig:kaon_fit_s_t}
\end{figure}

For the kaon, we plot the two-state fit data compared to fitted form factors for the vector in Fig.~\ref{fig:kaon_fit_v} and the scalar and tensor in Fig.~\ref{fig:kaon_fit_s_t} as done for the pion above. We find the fitted $F_\Gamma(0)$ to be independent of the fit range and the included data sets. There is some tension between the estimates of $M_S^{K^u}$ extracted from the rest frame and that of the boosted and combined frames. Similar behavior is observed in the tensor plot for the up-quark. 

As stated above, we choose the full $Q^2$ range and the combined frame for the fit parameters of each meson. However, for the radii calculations for the pion, we give results by constraining the fit up to $Q^2=0.55$ GeV$^2$. We also do not use the entire range for the kaon, constraining the fit up to $Q^2=1$ GeV$^2$. We report differences between radii extracted from differently constrained fits as systematic error. Table~\ref{tab:pion_radii} contains the fit parameters for the selected data sets, as well as the radii. Our results for $\rr^{\pi^u}_S$ are compatible with the ones obtained in Ref.~\cite{Gulpers:2013uca} from the connected contributions on an $N_f=2$ ${\cal O}(a)$-improved Wilson fermions ensemble at a pion mass of 280 MeV. Similar values are also obtained from a 310 MeV pion mass ensemble of $N_f=2+1$ overlap fermions~\cite{Kaneko:2010ru}. A sizeable logarithmic behavior in the pion mass is found in chiral perturbation theory~\cite{Gasser:1990bv,Bijnens:1998fm} which causes a rise in the radii. Therefore, at this stage, we do not attempt any comparison with the PDG value of $\rr^{\pi}_{V}$, as the ensemble we used is not at the physical value of the pion mass. Additionally, we observe that the extraction of the tensor radius is more sensitive to the fit range. We note that our results for $\rr^{K}_V$ are compatible with the ones of Ref.~\cite{Kaneko:2010ru} obtained from an $N_f=2+1$ ensemble of overlap fermions producing a pion mass of 310 MeV.

\begin{table}[h!]
\centering
\renewcommand{\arraystretch}{1.2}
\renewcommand{\tabcolsep}{3pt}
\resizebox{\textwidth}{!}{
\begin{tabular}{l c c c c c c c c c}
 \\ [-3ex]
 & $F_{S}(0)$ & $\,F_V(0)$ & $\,\kappa_T$ & $\,M_S$ & $\,M_V$ & $\,M_T$ & $\,\rr_{S}$ & $\,\rr_{V}$ & $\,\rr_{T}$\\
    \hline
$\pi^u$ & $1.165(6)(4)$ & $1.017(6)(6)$ & $0.376(5)(6)$ & $1.221(36)(60)$ & $0.832(8)(14)$ & $0.800(12)(29)$ & $0.232(22)(54)$ & $0.291(6)(36)$ & $0.461(44)(121)$\\
    \hline
$K^u$ & $1.093(8)(10)$ & $1.016(5)(11)$ & $0.844(9)(61)$ & $1.291(15)(40)$ & $0.822(5)(19)$ & $0.724(5)(59)$ & $0.149(3)(10)$ & $0.289(3)(13)$ & $0.382(4)(45)$\\
    \hline
$K^s$ & $1.158(7)(8)$ & $1.017(4)(11)$ & $0.717(5)(17)$ & $1.552(17)(46)$ & $1.000(6)(22)$ & $0.930(6)(37)$ & $0.103(2)(6)$ & $0.289(3)(13)$ & $0.250(3)(20)$\\
    \hline
\end{tabular}
}
\caption{The fit parameters for the pion and kaon form factors. The monopole masses are given in GeV and the radii in fm$^2$. The number in the first parentheses is the statistical uncertainty. The number in the second parentheses is the systematic error related to the fit range, and it is the the difference with the values using $Q^2_{\rm max}=1$ GeV$^2$ for the pion and $Q^2_{\rm max}=3$ GeV$^2$ for the kaon.}
\label{tab:pion_radii}
\vspace*{0.2cm}
\end{table}

\section{SU(3) flavor symmetry breaking}

The pion and kaon form factors are useful for studying SU(3) flavor symmetry breaking effects, which have been observed in nature in the charge radii of $\pi^{\pm}$ and $K^\pm$, as well as in $\pi^{0}$ and $K^0$. We examine the ratios $F^{\pi^u}/F^{K^u}$, $F^{\pi^u}/F^{K^s}$, and $F^{K^u}/F^{K^s}$ for the form factors to draw conclusions on these effects. Here, we only show results for the vector case. Since the value of $Q^2$ depends on the mass of the meson, we use the fitted values of the form factors in these ratios. We are also interested in the effects of excited-state contamination on the ratios, so we use the parameterizations on individual plateau values as well as the two-state fit.

\begin{figure}[h!]
    \centering
    \includegraphics[scale=0.23]{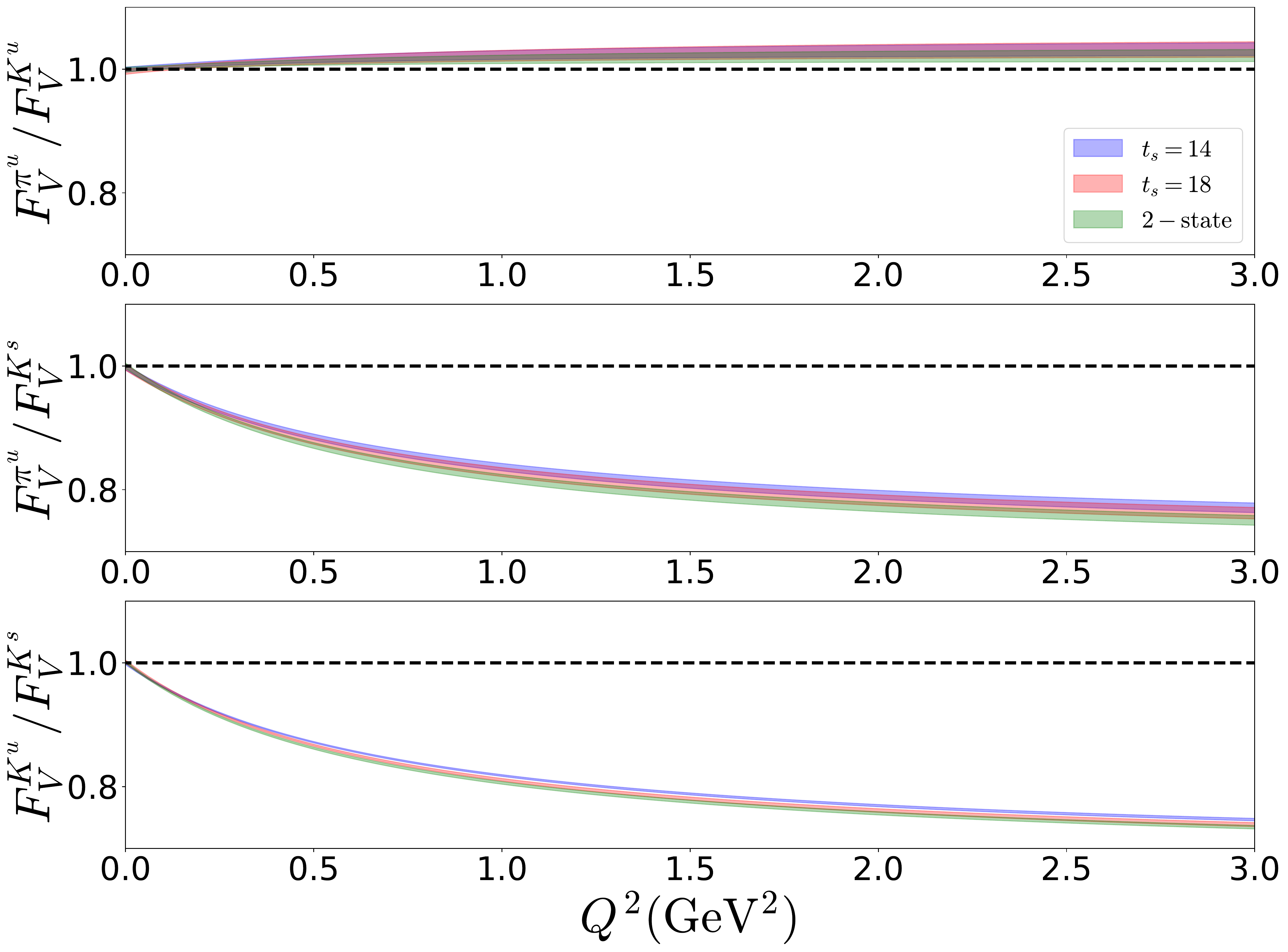}
    \caption{The ratio $F_V^{\pi^u}/F_V^{K^u}$ (top), $F_V^{\pi^u}/F_V^{K^s}$ (center), and $F_V^{K^u}/F_V^{K^s}$ (bottom) for the vector form factor as a function of $Q^2$ using the results obtained from both frames. The results for $t_s/a=14,\,18$ and the two-state fit are shown with blue, red and green bands, respectively.}
    \label{fig:SU3_vector}
\end{figure}

In Fig.~\ref{fig:SU3_vector} we show the ratios described above, and find that the excited-state contamination is much more suppressed than the individual form factors. Notably, the ratio $F^{\pi^u}/F^{K^u}$ has little $Q^2$ dependence and is nearly 1. Due to the similarity in the up-quark component of the pion and the kaon, we find the up-quark contribution to be about $80\%$ of that of the strange-quark for the ratios $F^{\pi^u}/F^{K^s}$ and $F^{K^u}/F^{K^s}$.

\section{Summary}

We present a calculation in lattice QCD of the scalar, vector, and tensor form factors for the pion and kaon obtained on an $N_f=2+1+1$ ensemble of twisted mass fermions with clover improvement that corresponds to 265 MeV pion mass and 530 MeV kaon mass. We renormalize the scalar and tensor form factors non-perturbatively and give the results in the $\overline{\rm MS}$ scheme at a scale of $2$ GeV. The vector form factor does not need renormalization as we use the conserved vector operator. 

We utilize two kinematic setups to obtain the form factor: the rest frame and a momentum-boosted frame of 0.72 GeV ($\mathbf{p'}=\frac{2\pi}{L}(\pm1,\pm1,\pm1)$). We use a factor of 50 more statistics in the boosted frame compared to the rest frame in order to control statistical uncertainties. We extract the form factors up to $Q^2=2.5$ GeV$^2$ for the pion and up to $Q^2=3$GeV$^2$ for the kaon. Due to frame independence, we are able to combine the data of the rest and boosted frames. We find excellent agreement between the two frames for the vector form factors of both mesons, as well as for the strange-quark contributions for the scalar and tensor form factors of the kaon. We find good agreement in the small-$Q^2$ region for the up-quark part of the pion and kaon scalar and tensor form factors with deviations in the slope from $0.25-0.5$ GeV$^2$ for the pion and $0.35-1$ GeV$^2$. This indicates systematic uncertainties, such as cutoff effects.

We give final results for the form factors using the two-state fits and parameterize their $Q^2$ dependence using a monopole fit. This leads to the scalar, vector, and tensor monopole masses and the corresponding radii. We also extract the tensor anomalous magnetic moment, $\kappa_T$, which can only be obtained from fits on the tensor form factor data. In the study of the sensitivity of the extracted parameters on the fit range of $Q^2$ and the included frames, we find some tension in the scalar and tensor monopole masses and radii based on the data sets included in the fit. For the pion radii we use all data up to $Q^2=0.5$ GeV$^2$ and up to $Q^2=1$ GeV for the kaon. We provide a systematic error by varying the fit range.

We address SU(3) flavor symmetry breaking effects by comparing the parameterized form factors for the pion and kaon. We find that excited states ar suppressed in these ratios. Additionally, we find mild $Q^2$ dependence in the $F^{\pi^u}/F^{K^u}$ ratio for all operators. For the $F^{\pi^u}/F^{K^s}$ and $F^{K^u}/F^{K^s}$ cases we find SU(3) flavor symmetry breaking effects up to $20\%$. 

\section{Acknowledgements}

We would like to thank all members of ETMC for a very constructive and enjoyable collaboration. M.C. thanks Martin Hoferichter for interesting discussions on Ref.~\cite{Hoferichter:2018zwu}.  
M.C. and J.D. acknowledge financial support by the U.S. Department of Energy Early Career Award under Grant No.\ DE-SC0020405. 
K.H. is financially supported by the Cyprus Research Promotion foundation under contract number POST-DOC/0718/0100, CULTURE/AWARD-YR/0220 and EuroCC project funded by the Deputy Ministry of Research, Innovation and Digital Policy, the Cyprus The Research and Innovation Foundation and
the European High-Performance Computing Joint Undertaking (JU) under grant agreement No. 951732. The JU received
support from the European Union’s Horizon 2020 research and innovation programme.
S.B. is supported by the H2020 project PRACE 6-IP (grant agreement No 82376) and the EuroCC project (grant agreement No. 951732).
C.L. is supported by the Argonne National Laboratory with a research subcontract with Temple University.
A.V is supported by the U.S. National Science Foundation under Grants No. PHY17-19626 and PHY20-13064.
This work was in part supported by the U.S. Department of Energy, Office of Science, Office of Nuclear Physics, contract no.~DE-AC02-06CH11357 and the European Joint Doctorate program STIMULATE funded from the European Union’s Horizon 2020 research and innovation programme under grant agreement No 765048.
This work used computational resources from Extreme Science and Engineering Discovery Environment (XSEDE), which is supported by National Science Foundation grant number TG-PHY170022. 
It also includes calculations carried out on the HPC resources of Temple University, supported in part by the National Science Foundation through major research instrumentation grant number 1625061 and by the US Army Research Laboratory under contract number W911NF-16-2-0189. 

\bibliographystyle{ieeetr}
\bibliography{references}

%

\end{document}